\documentclass[10pt, final]{IEEEtran}

\usepackage{graphicx}
\usepackage{dcolumn}
\usepackage{bm}

\usepackage[english]{babel}
\usepackage{amssymb}
\usepackage{amsmath}
\usepackage{bbm}
\usepackage{nicefrac}
\usepackage{braket}
\usepackage{latexsym}
\usepackage{soul}
\usepackage{braket}
\usepackage{tikz}
\usepackage{cancel}
\usepackage{caption}
\usepackage{subcaption}
\usepackage{physics}

\usepackage{cite}

\usepackage{hyperref}
\hypersetup{
 colorlinks=true,
 linkcolor=blue,
 filecolor=magenta, 
 urlcolor=cyan,
}
 
\begin{document}

\title{Quantum hobbit routing: Annealer implementation of generalized Travelling Salesperson Problem}

\author{
Iñigo Pérez Delgado{$^1$}, Beatriz García Markaida{$^2$}, Aitor Moreno Fdez. de Leceta{$^3$},
Jon Ander Ochoa Uriarte{$^4$}

\medskip

{$^1$}Qi3B Ibermatica, Parque Tecnológico de Bizkaia, Ibaizabal Bidea, Edif. 501-A, 48160 Derio, Spain \\
Email: i.perez.delgado@ibermatica.com

{$^2$}Qi3B Ibermatica, Parque Tecnológico de Bizkaia, Ibaizabal Bidea, Edif. 501-A, 48160 Derio, Spain \\
Email: b.garcia@ibermatica.com

{$^3$}Qi3B Ibermatica, Parque Tecnológico de Alava, Leonardo Da Vinci, 01510 Miano, Spain \\
Email: ai.moreno@ibermatica.com

{$^4$}Economic and Commercial Office of the Embassy of Spain, Singapore \\
Email: jon.ochoa@comercio.mineco.es
}

\maketitle

\begin{abstract}
In this paper, we present an implementation of a Job Selection Problem (JSP) --- a generalization of the well-known Travelling Salesperson Problem (TSP)--- of $N=9$ jobs on its Quadratic Unconstrained Binary Optimization (QUBO) form, using $\mathcal{O}(N)$ qubits on DWave's Advantage$\_$system4.1 quantum annealing device. The best known quantum algorithm for TSP to date uses $\mathcal{O}(N^2)$ qubits. A solution is found using the quantum method. However, since hardware is not yet able to compensate the increase in search-space size, no present overall advantage is achieved when comparing the quantum results with either exhaustive or equiprobably sampled classical solutions of the problem.
\end{abstract}


\section{\label{sec:intro}Introduction}

The usage of quantum devices as a mean of information transfer beyond classical physics was born as a mental experiment by Wiesner \cite{Wiesner1983but68} in 1968 \cite{Mor2014wiesner,InterviewBennett}. However, that paper was not published until the early 80s, after the reignition of the interest on the topic caused by the publications of Rabin \cite{Rabin1981oblivious}, Diek \cite{Dieks1982nocloning}, and Wootters and Zurek \cite{Wootters1982nocloning}. Paralelly, along with the demonstration of the Turing completeness of `quantum states' by Benioff \cite{Benioff1980qTuring} {---which would now be called `qubits' \cite{schumacher1995qubit}---} Manin \cite{manin1980computable} and Feynman \cite{feynman1982computer} started to think of quantum devices as computational machines: if quantum systems are hard to simulate by classical computers, and analog quantum systems can be used to mimic the evolution of other quantum systems, then it is clear that quantum computers have capabilities beyond those of classical devices. {It is only now, more than half a century after Wiesner's paper, that real quantum devices are large and stable enough to attempt a rich variety of tasks. The current state of quantum computing, referred to as the Noisy Intermediate-Scale Quantum (NISQ) era, is however still far from fault-tolerant systems, and thus noise heavily limits the size of the problems solvable by modern quantum devices. The chosen problem, called Job Selection Problem (JSP), lies in this frontier, where we are able to obtain the solution with the chosen quantum device ---the Advantage$\_$system4.1 DWave quantum annealer--- but obtaining no present improvement when compared with classical solutions. However, we do improve on previous similar quantum algorithms, reducing the number of required qubits from $\mathcal{O}(N^2)$ to just $\mathcal{O}(N)$ thanks to a specifically chosen formulation.}

The JSP is a generalization of the widely known Travelling Salesperson Problem (TSP). {In the JSP, the traveller needs not only to optimize the route through a number of nodes like in a TSP, but has to first select which of all the possible places are they going to visit, since only a limited amount of some resource is available for the travel and thus typically it is not possible to visit all nodes. In our problem this limiting factor is time, but it could be fuel, distance, or another resource. Then, instead of all $N$ nodes being visited in $N$ timesteps, only $\xi$ nodes are visited out of all $N$ possible ones, using only $\xi$ timesteps.}

{Our JSP-focused formulation allows us to use only $\mathcal{O}(\xi N)$ qubits instead of the $\mathcal{O}(N^3)$ of the native formulation of the TSP \cite{Papalitsas2019native}, the $\mathcal{O}(N^2\log_2N)$ of MTZ formulation \cite{Miller1960MTZ} or the $\mathcal{O}(N^2)$ of GPS formulation \cite{Gonzalez-Bermejo2022GPS}, all of them designed for the TSP. For our JSP formulation, in the $\xi(N)=N$ case where we recover the TSP, $\mathcal{O}(\xi N)=\mathcal{O}(N^2)$, which matches the GPS formulation. However, there are several instances where $\xi$ is not a function of $N$. For example, a delivery drone could have 12 hours of battery independently of the number of packages waiting to be delivered. In those cases, $\mathcal{O}(\xi N)$ is really $\mathcal{O}(N)$. To the best of our knowledge no quantum solution of the JSP has been published yet, so this is the first time this formulation has been used in this context.}

{In this paper we will first illustrate the quantum process with the simple example of Sec. \ref{sec:advantage}, which will help the reader understand both where the advantage quantum annealing provides comes from and where the problems of the technique may be found. In Sec. \ref{sec:problem} the exact instance of JSP will be explained, and some narrative context will be provided in order to improve the comprehensibility of the problem. In Sec. \ref{sec:hamil} the details of the hamiltonian of the model will be explained, and in Sec. \ref{sec:constants} the selection of the values of its weights will be justified. Once the hamiltonian has been presented, the problem will first be solved classically in Sec. \ref{sec:classical}, both random and exhaustively. As a sanity check, in \ref{sec:confirmation} those results will be used to confirm the validity of the proposed hamiltonian. Lastly, in Sec. \ref{sec:results} the results of the quantum processing will be presented, and the extent of the success of the method will be shown. The conclusions of the work appear in Sec. \ref{sec:conclusions}.}

We have chosen to set the problem in the world of Tolkien's The Lord of the Rings trilogy, in order to provide an easy-to-follow example in a fictional yet hopefully familiar world.

\section{\label{sec:advantage}The value of quantum}

The DWave annealing protocol initiates each qubit to the uniform superposition $(\ket{0}+\ket{1})/\sqrt{2}$, which combined equals to a uniform superposition of all of the $2^N$ individual eigenstates of the $N$-qubit base. Then, the state is left to cool down, which transfers the probability amplitude of high-energy states into low-energy ones. This means that, when measuring the system, it will collapse with higher probability to lower-energy states. If the hamiltonian of the system is constructed to inversely correlate the energy of the states and the quality of the solutions they represent, in order to find the explicit form of the solution we will just need to perform the quantum experiment a number of runs inversely proportional to the probability of measuring the lowest-energy state (or states, in case of degeneration).

\begin{figure}[h]
  \includegraphics[width=\linewidth]{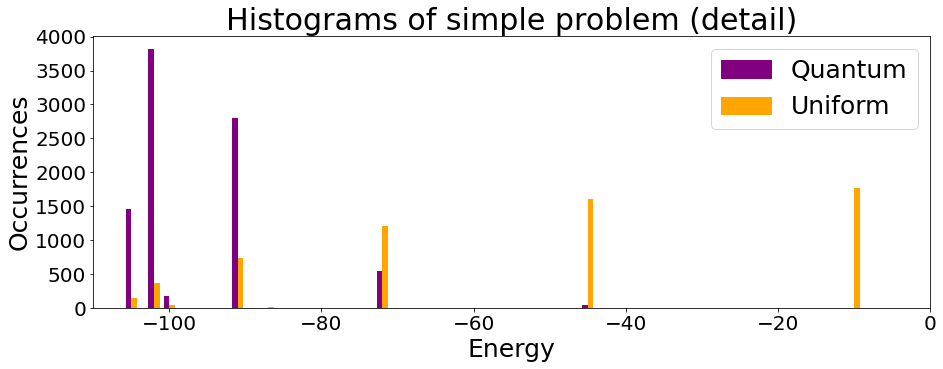}
  \caption{Results of the quantum annealer ---in purple---, albeit random in nature, show an increase of the measuring probability of low-energy answers when compared with a sample built choosing each variable $x_i\in \{0,1\}$ equiprobably ---in orange.}
  \label{fig:equiprobable}
\end{figure}

In order to visualize the effect of the cooling down of the system, we have solved a dummy problem with the DWave annealer. In this problem, of $N=20$ binary variables, the solutions are those states with five of the variables equal to 1 and the remaining fifteen equal to 0. Results of the experiment are shown in Fig. \ref{fig:equiprobable}.

The increase in the measuring probability of the ground state is crucial, since in several kinds of problems the number of answers corresponding to high-energy states is far greater than the number of optimal answers. In our simple $N=20$ example, only $20 \choose{5}$ $=15504$ answers have the minimum energy from all $2^{20}=1048576$ possible states, and it is easy to see how quickly this difference escalates with big values of $N$. 

\section{\label{sec:problem}Description of the problem}

{Years after the events of the trilogy of The Lord of the Rings, Samwise Gamgee decides he wants to go on another trip around the Middle earth. In total, his list is composed of $N=9$ places, listed in Tab. \ref{tab:priorities}. However, he wants to be back home by the 1st of May, which gives him $t_{MAX}=100$ days for the trip. This means he will need to select only some of the places of his list, leaving some others out.}

He procceeds to assign a priority to each of the places he wants to visit on Tab. \ref{tab:priorities}. The second thing Sam needs to know in order to plan his trip is the distances between all these $N$ places. {He tabulates them on Tab. \ref{tab:distances}. Lastly, he estimates how much time he will spend on each of the destinations, and adds them to Tab. \ref{tab:priorities}.}

\begin{table}[h]
    \centering
     \caption{The visiting priorities of each place and the expected duration of the visit. According to the table, for example, Sam would enjoy equally visiting Rivendel $(p_{tot}=40)$ and visiting both Isengard and Tharbad $(p_{tot}=35+5)$.}
    \begin{tabular}{|c|c|c|}
    \hline
    PLACE $i$ & PRIORITY  $p_i$ & VISIT TIME $t_i$\\
    \hline
    Bree     & 15 & 3 days\\
    Edoras     & 150 & 5 days\\
    Isengard     & 35 & 4 days\\
    Lórien     & 75 & 4 days\\
    Minas Tirith     & 170 & 7 days\\
    Pelargir     & 50 & 3 days\\
    Rivendel & 40 & 5 days\\
    Tharbad    & 5 &2 days\\
    Valle     & 15 & 4 days\\
    \hline
    \end{tabular}
    \label{tab:priorities}
\end{table}


{Lastly, an average travelling speed is needed to convert disctances into time. Sam's pony moves at $v=9.6$ leagues a day \cite{TLOTRRPG}.}

\section{\label{sec:hamil}The hamiltonian} 

Let us describe the algorithm Sam could use if he had access to a quantum annealing device. The core idea of quantum annealer problems is to write a hamiltomian whose state of minimum energy is the state that best fulfills our criteria. That state, in our algorithm, will be described by variables labeled as $x_{i,s}$ and stored each in one qubit. If the $i$th location has been visited in $s$th place, then the variable $x_{i,s}$ will be equal to one. Else, $x_{i,s}=0$, since we will write this problem in its Quadratic Unconstrained Binary Optimization (QUBO) form. As $i\in\{1,N\}$ and $s\in\{1,\xi \}$, the number of qubits needed is $\xi  N$. For problems where each place is only visitable once, we have that $N\geq \xi $. Moreover, it is also easy to find problems where $N>> \xi $, so the usefullness of our algorithm is apparent. However, the reduction in the number of qubits comes with an increase of the number of problems to be solved: the method we propose is to solve the problem for $2\Delta$ different values of $\xi $ centered around the number of steps $s_{cl}$ of some classically calculated solution. This solution does not need to be a very fine solution either, because it is only intended to give us a characteristic size for the problem. This increase in the number of steps is not an issue either, since $\Delta<<N$ and thus solving $2\Delta$ problems of $N \left(s_{cl}\pm\Delta\right)$ variables is at worst equivallent to solving one problem of $\mathcal{O}(N^2)$ variables. As an example of the relative size of those numbers, one realistic problem could have $N=1000$, $s_{cl}=20$ and $\Delta=5$, which would mean solving 10 problems of sizes between 15,000 and 25,000 variables instead of one of size 1,000,000.

For clarity purposes, we will divide our Hamiltonian in two parts: the `natural' hamiltonian $H^0$ and the `restriction' hamiltonian $H^R$. Then, the total hamiltonian will be a sum of both: 

\begin{equation}
    H=H^{0}+H^{R}
\end{equation}

\subsection{Natural hamiltonian}

With the `natural' part of the hamiltonian we account for all natural causes of energy of our analog system: we want to maximize the total solved priority, so we will add a negative term $H_p$ proportional to it. However, we want to do it in the minimum time possible, so another two positive terms will be added too, proportional to the times spent travelling to ($H_{tt}$) and visiting ($H_{vt}$) each place. 

\begin{table*}[tp]
    \centering
    \caption{The distances $d_{ij}$ between different places of the middle earth, in leagues \cite{TLOTRRPG, TLOTRRPG_errata}.}
    \begin{tabular}{|c| c c c c c c c c c c|}
    \hline
     & Bree & Edoras &Isengard  & Hobbiton  & Lórien & Minas Tirith     &Pelargir &Rivendel & Tharbad & Valle\\
    \hline
    Bree    & - & 200 & 150 & 40 & 140 & 285 & 315 & 100 & 67 & 225 \\
    Edoras   &  200 & - & 48 & 225 & 100 & 102 & 117 & 172 & 133 & 235  \\
    Isengard   &    150 & 48 & - & 175 & 83 & 150 & 163 & 135 & 83 & 225  \\
    Hobbiton & 40 & 225 & 175 & - & 183 & 321 & 342 & 167 & 90 & 270 \\
    Lórien    & 140 & 100 & 83 & 183 & - & 158 & 192 & 77 & 100 & 145  \\
    M. Tirith & 285 & 102 & 150 & 321 & 158 & - & 43 & 200 & 229 & 245  \\
    Pelargir      & 315 & 117 & 163 & 342 & 192 & 43 & - & 243 & 252 & 290 \\
    Rivendel  & 100 & 172 & 135 & 167 & 77 & 200 & 243 & - & 100 & 125 \\
    Tharbad     & 67 & 133 & 83 & 90 & 100 & 229 & 252 & 100 & - & 220 \\
    Valle      & 225 & 235 & 225 & 270 & 145 & 245 & 290 & 125 & 220 & -\\
    \hline
    \end{tabular}

    \label{tab:distances}
\end{table*}

\begin{equation}
H^0\equiv H_p+H_{tt}+H_{vt}\;.
\end{equation}

Remember that we want the best fitting state to be the ground state of the hamiltonian, that is, the state with the lowest energy, so if a term is negative it will actually represent a preferred state. 

\subsubsection{Priority term}\label{sssec:priority}

The first term of our hamiltonian will be a negative term that accounts for the total obtained priority:

\begin{equation}
    H_p=-c_p\sum_{i,s} p_i x_{i,s}\;,
\end{equation}

where $c_p>0$ is a proportionality constant that will help us balance the contribution of each term of the hamiltonian. The more places we visit (the more $x_{i,s}\in\{0,1\}$ are equal to one) more priority terms $p_i$ will be added, making the hamiltonian more negative.

\subsubsection{Travel time term}

The second term will be a positive term that accounts for the travelling time:

\begin{equation}
\begin{split}
    H_{tt}&=c_{tt}\Bigg[\sum_{i j,s} \frac{d_{ij}}{v} x_{i,s}\,x_{j,s+1} \\
    + &\sum_{i} \frac{d_{0i}}{v} x_{i,1} + \sum_{i} \frac{d_{i0}}{v} x_{i,\xi }\Bigg]\;,
\end{split}
\end{equation}

where $c_{tt}>0$ is another proportionality constant. The first of the sums of $H_{tt}$ accounts for the time spent travelling between visited places: if (and only if) a place $i$ is visited at step $s$ (that is, $x_{i,s}=1$) and then place $j$ is visited at step $s+1$ ($x_{j,s+1}=1$) then we add a positive energy term. This term is proportional to the time spent moving from place $i$ to place $j$ which, assuming a constant speed $v$, is proportional to the distance $d_{ij}$ that separates both places.

The second sum accounts for the time travelling from home, denoted as $i=0$, to the place visited in the first step, and is proportional to the distance $d_{0i}$. Similarly, the third sum counts the time spent travelling from the last visited place back home, and is proportional to the distance $d_{i0}$. Note that in a symmetrically distanced problem such as this  $d_{ij}=d_{ji}$ and the distinction is just for notation clarity.

\subsubsection{Visit time term}

We will add a third term, also positive, accounting for the visit time, with its own proportionality constant $c_{v}>0$:

\begin{equation}
    H_{vt}=c_{vt}\sum_{i,s} t_i x_{i,s}\;.
\end{equation}

If place $i$ is visited at timestep $s$, $x_{i,s}=1$ and a term proportional to $t_i$ will be added.

\subsection{Restriction hamiltonian}

When using only the natural part of the hamiltonian, however, some problems arise. For example, as each place $i$ is represented $\xi $ times by variables \{$x_{i,1}$, $x_{i,2}$, $x_{i,...}$\}, nothing stops the hamiltonian to `visit' a high-priority place $i_0$ multiple times, getting to add $-c_p p_{i_0}$ on each of the timesteps and thus lowering the total energy to its minimum. In fact, nothing stops the hamiltonian to turn all $x_{i,s}$ to one, visiting all the places all the time. 

It would seem that a QUBO method, unconstrained by definition, should not be able to take into account those requisites. However there is a simple way to implement them in the hamiltonian. Those terms are called the `restriction hamiltonian', $H^{R}$. 

\subsubsection{`One place per timestep' term}\label{sssec:ops}

In order to have exactly one place visited on each step $s$, we want all $x_{i,s}$ to be zero $\forall i$ except for one. The following equations will then be fulfilled:

\begin{equation}
    \forall s, \quad \sum_i x_{i,s}=1 \;.
    \label{eq:ops}
\end{equation}

If the criteria are not met, the energy should go up. Hence, to ensure our nodes are visited in a one-per-step fashion, we will add the following sum of terms to our hamiltonian: 

\begin{equation}
    H_{ops} = \lambda_{ops}\sum_s \left(\sum_i x_{i,s}- 1\right)^2 \;,
\end{equation}

where $\lambda_{ops}>0$ is a proportionality constant, similar in nature to the $c$ constants of the terms of $H^0$ but much bigger. $\lambda$ constants are sometimes called Lagrange multipliers.  If each of the Eqs. \ref{eq:ops} is fulfilled, $H_{ops}=0$ and no penalty is imposed. Else, $H_{ops}>0$ and, as $\lambda_{ops}$ is big enough, that is translated into a penalty unassumable by the system, which means only solutions which fulfill Eqs. \ref{eq:ops} will appear as feasible.

\subsubsection{`Each place is visited once at most' term}

In a similar fashion to Sec. \ref{sssec:ops}, in order to ensure no place is visited more than once, a penalty term will be added. In this case we want to fulfill 

\begin{equation}
    \forall i, \quad \sum_s x_{i,s}\leq1 \;,
    \label{eq:oam}
\end{equation}

which is not a set of equations but of inequalities. Usually this means extra dummy variables are required, which means more qubits, but in this special case where only two values are permitted (a place can be visited either zero times or one time) we can write the term $H_{oam}$ (\textit{`once at most'}) as follows:

\begin{equation}
    H_{oam} = \lambda_{oam}\sum_i \left(\sum_s x_{i,s}- 0.5\right)^2\;.
\end{equation}

If a place $i$ is visited zero or one times the penalty is not zero, but it is the minimum value it can take: $\lambda_{oam}(\pm0.5)^2$. Visiting any other number of times \{2,3,4,...\} will make the penalty term higher than that \{$\lambda_{oam}(1.5)^2$, $\lambda_{oam}(2.5)^2$, $\lambda_{oam}(3.5)^2$, ...\}.

\medskip

\section{\label{sec:encoding}Hamiltonian encoding}

If we take a look at our hamiltonian, all the individual terms are either constant, in which case do not affect the localization of the state of minimum energy, or proportional to one or to two $x_{i,s}$ variables. However, as we are dealing with a QUBO prooblem, $x_{i,s}\in\{0,1\}$ by definition, and we have that $x_{i,s}=\left(x_{i,s}\right)^2$. Then, effectively, all the terms of the hamiltonian we need to take into account are proportional to some second-order term $x_{i,s}\,x_{j,s'}$, where diagonal $x^2_{i,s}$ terms will correspond to originally $x_{i,s}$ first-order terms. Depending of the exact syntaxis of the specific annealer controller, we will store the coefficients of these terms either in an $N\xi \times N\xi $ matrix or, as in our case, in a two-keyed dictionary with that same number of entries ($N\xi $ for each key). In any case the storage item will be denoted as $Q$. 

\section{\label{sec:constants}Selection of constants}

As mentioned in Sec. \ref{sec:hamil}, coefficients $c_p$, $c_{tt}$, $c_{vt}$, $\lambda_{ops}$ and $\lambda_{oam}$  have to be chosen in a proportion such that the minimum of the energy corresponds to a maximum of the total solved priority. {There are multiple combinations of values that satisfy that minimal requirement. However, some of them will give a higher probability of finding a solution than others. As explained in each case, approximated values for each of the constants have been calculated, and trial-and-error empirical adjustments have been performed in order to complement them.}

These are the chosen constants:

\begin{itemize}
    \item $c_p=0.1$, so that the typical energies of the system are of the order of 10s and not 100s. {This is mostly an aesthetic choice, but has been left intentionally in the manuscript to underline that only the proportion between constants is relevant.}
    \item $c_{tt}=c_{vt}=c_p(p_{GUESS}/t_{MAX})$. Here, $p_{GUESS}$ is a rough estimate of the priority of our optimal answer, which gives the hamiltonian a feel of the characteristic size of the problem. In this paper we have used $p_{GUESS}=500$, but $p_{GUESS}=400$ or $p_{GUESS}=600$ would have equally worked. Since only the proportion between terms matter, we took priority terms as our reference. Time terms are always penalties, so it makes sense to decrease their coefficients when $t_{MAX}$ is big and so time is a less scarce resource.
    \item $\lambda_{ops}=300c_p$ and $\lambda_{oam}=200c_p$, since penalties have to be big enough to be unassumable: we impose that $\forall i\; |\lambda_{ops}|, |\lambda_{oam}| > |c_p p_i|$. {$\lambda_{ops}$ is slightly bigger because, empirically, setting it at $200c_p$ still did not avoid a significant number of violations of its intended restriction due to the non-ideal nature of harware.}
\end{itemize}

\section{\label{sec:classical}Classical resolutions}
\subsection{\label{sec:exhaustive}Exhaustive resolution}

The simplest way of obtaining the optimal route between those which fulfil our requisites is to simply check every possible route for each $\xi$, filtering out those who do not met the criteria, and then looking for the best answer between all the remaining routes. The best answers will be those with a maximum total priority $p_{tot}^\xi$, calculated by the sum of the priorities of each of the visited places. $n_o^\xi$ is the number of separate answers that, for each $\xi$, share the same $p_{tot}^\xi$. Due to the distance symmetry of the problem, $n_o^\xi$ has to be even, since forwards and backwards versions of the same route are counted as separate answers. We then obtain the $p_{tot}^G$ priority of the global optimum or optima by choosing the largest of the $p_{tot}^\xi$ priorities. $n_o^G$ is the number of answers that have priority $p_{tot}^G$. 

Doing so for our problem, one finds out there are $n_o^G=12$ equivalent globally optimal routes, all of them with $\xi =6$, and with $p_{tot}^G=495$. This is shown in Tab. \ref{tab:EX_results}. The runtimes required to check all routes for each $\xi$, denoted as $T_{run}$, also appear on the table.

\begin{table}[h]
    \centering
        \caption{Number and priority of optimal routes for different values of $\xi $, obtained through exhaustive search. Runtimes of each search are also shown. Forwards and backwards versions of the same route are counted as separate answers here.}
    \begin{tabular}{|c|c|c|c|}
    \hline
    $\;\;$ $\xi$ $\;\;$&$\;\;$ $p_{tot}^\xi$ $\;\;$&$\;\;$ $n_o^{\xi }$  $\;\;$&$\;\;$ $T_{run}^\xi$ $\;\;$\\
    \hline
    
    4 & 445& 8  & 0.09s \\
    5 & 480& 4  & 0.39s  \\
    6  & 495& 12  & 1.61s\\
    7  & 345& 2  & 5.53s\\
    8  & - & 0  & 13.40s\\
    \hline
    Global & 12 & 495 & 21.02s\\
    \hline
    \end{tabular}
    \label{tab:EX_results}
\end{table}

However, the runtime escalates quickly not only with the size of the problem, but with the number of steps, proportionally to the  number $f$ of possible routes 

\begin{equation}
    f(\xi )=\frac{N!}{(N-\xi )!}
    \label{eq:f}
\end{equation}

which, when $\xi <<N$, can be approximated by
\begin{equation}
    f(\xi )\approx N^{\xi }\;.
\end{equation}

\begin{figure}[h]
  \includegraphics[width=\linewidth]{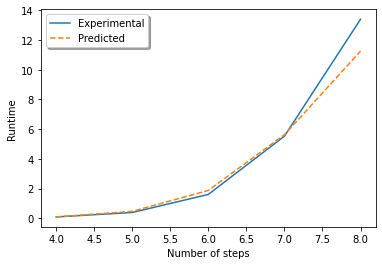}
  \caption{Predicted and experimental runtimes for the exhaustive classical search of optimal routes for our $N=9$ sized problem.}
  \label{fig:Runtime_comparison}
\end{figure}

The comparison between the real runtime and the runtime calculated with Eq. \ref{eq:f} is shown in Fig.\ref{fig:Runtime_comparison}.

\subsection{\label{sec:complete}Random sample resolution}

For problems with $N>>\xi $ the exhaustive method escalates quickly with both variables. Moreover, exhaustive search is not what quantum annealers do, since they work with a fixed number of probabilistic runs. Adopting that same approach with a classical computer allows us to drastically reduce the runtime of the search for optimal routes, at the expense of having to deal with the intrinsic uncertainty of random processes. 

Let $f(\xi )$ be the number of possible different routes for a given $\xi $. Then, knowing from Tab. \ref{tab:EX_results} the number of those routes which are optimal, we can calculate the $P^\xi$ probability that, for one randomly chosen route of a certain $\xi $, that route is optimal. Then, for $r$ runs, we can find the expected number of found optimal answers to be

\begin{equation}
    \langle n_f^\xi \rangle=rP^\xi = \frac {r\;n_o^\xi}{f(\xi)}=\frac {r\;n_o^\xi (N-\xi)!}{N!}\;.
\end{equation}

Performing the experiment  a number $r=10000$ of runs for each $\xi$, we obtain the results shown in Tab. \ref{tab:RS_results}.

\begin{table}[h]
    \centering
        \caption{The number of times a randomly chosen route of $\xi$ steps was found to have the optimal priority $p_{tot}^\xi$, as read from Tab. \ref{tab:EX_results}, out of $r=10000$ runs. The combined runtime $T_{run}^\xi$ of those $r$ runs is also shown for each $\xi$.}
    \begin{tabular}{|c|c|c|c|}
    \hline
    $\;\;$ $\xi $ $\;\;$ & $\;\;$ $n_f^\xi$ $\;\;$ & $\;\;$ $\langle n_f^\xi \rangle$ $\;\;$ &$\;\;$ $T_{run}^\xi$ $\;\;$\\
    \hline
    
    4 & 29 & 26.46 &  0.23s \\
    5 & 3 & 2.65 & 0.28s\\
    6  & 5 & 1.98 & 0.34s\\
    7  & 0 & 0.11 & 0.39s\\
    8  & 0 & 0 & 0.47s\\
    \hline
    \end{tabular}
    \label{tab:RS_results}
\end{table}

Although $T_{run}^\xi$ escalates much better in Tab. \ref{tab:RS_results} than in Tab. \ref{tab:EX_results}, the use of a fixed number of $r$ runs is excessive for low-$\xi$ cases, where $\langle n_f^\xi \rangle>n_o^\xi $, and not enough for high-$\xi$ cases where $\langle n_f^\xi \rangle<1$.

\section{\label{sec:confirmation}Hamiltonian confirmation}

\begin{figure}[h]
  \includegraphics[width=\linewidth]{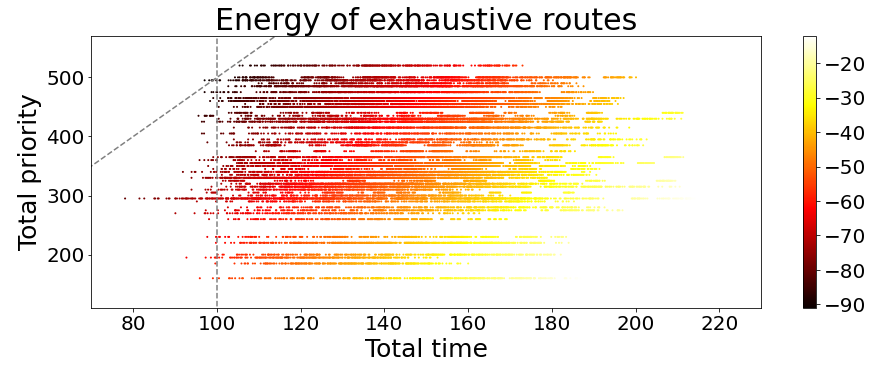}
  \caption{Energies of all the possible $\xi=6$ routes which visit each place once at most. The energy gradient is perpendicular to the $H^0=0$ diagonal line. {The $t=t_{MAX}$ vertical line is also shown: only results on the left side of the line are valid solutions.} }
  \label{fig:exhaustive}
\end{figure}

\begin{figure}[h]
  \includegraphics[width=\linewidth]{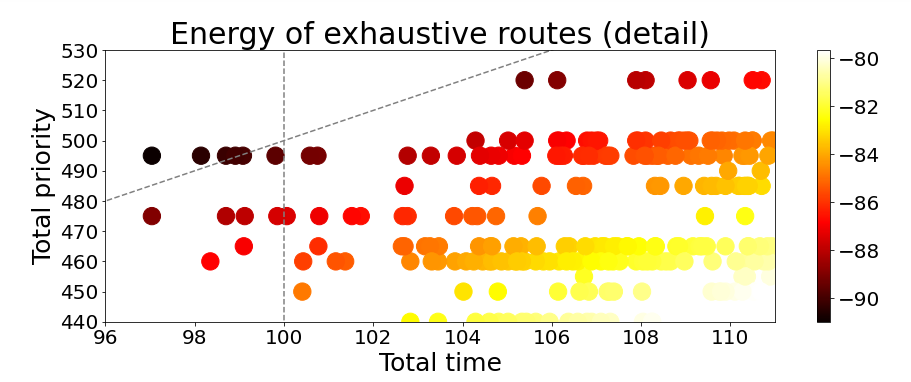}
  \caption{Detail of Fig. \ref{fig:exhaustive}. {There are 6 distinct  routes with $p=495$ which are legal ($t<100$), which represent our optimal answers. They have the minimum energies of the whole set.} Moreover, they are well-ordered by our secondary criteria, total time. {$H^0=0$ diagonal line is also shown.}}
  \label{fig:Exhaustive_detail}
\end{figure}

Once that we know that the optimum answer is a route with $\xi=6$ and $p=495$, we can check whether the hamiltonian devised in Sec. \ref{sec:hamil} has that answer as its lowest-energy state. This, as the classical resolutions of \ref{sec:classical}, would not be part of the proposed method and is used here only for comparison and confirmation purposes. Plotting all the $\xi=6$ routes as a function of their $p_ {tot}$ total priority and $t_{tot}$ time needed to complete them, along with the $t=t_{MAX}$ vertical line and the $H^0=0$ diagonal line where $|H_p|=|H_{tt}+H_{vt}|$, gives us the figures \ref{fig:exhaustive} and \ref{fig:Exhaustive_detail}.

It is enough to be the lowest-energy routes amongst the exhaustive search, since any change that decreases energy by adding or repeating a high-priority place to the route will be affected by a penalisation even greater, since $\forall i\; |\lambda_{ops}|, |\lambda_{oam}| > |c_p p_i|$.

\begin{figure}[h]
  \includegraphics[width=\linewidth]{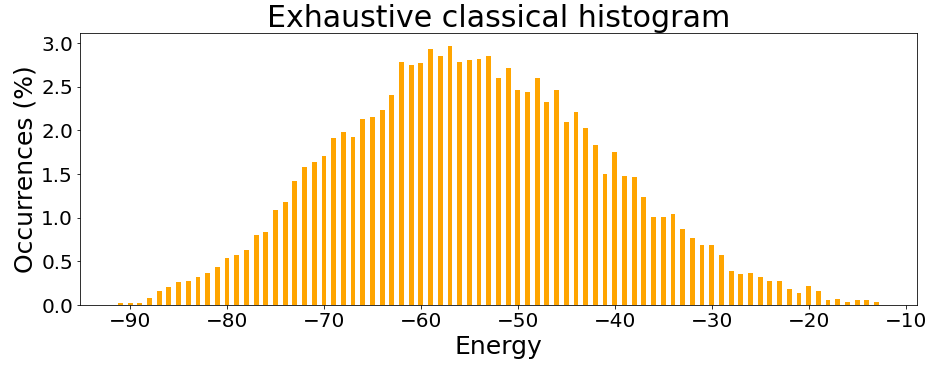}
  \caption{Distribution of energies of all the possible $\xi=6$ routes which visit each place once at most. The optimal $p=495$ answers correspond to the lower-energy end of the histogram.}
  \label{fig:exhaustive histo}
\end{figure}

The distribution of results by energy is shown in Fig. \ref{fig:exhaustive histo}. As in Fig. \ref{fig:equiprobable}, the classical distribution follows the bell distribution expected from a combinatorial problem like this.

\section{\label{sec:results}Results}

Performing the experiment three times in the quantum annealer Advantage$\_$system4.1, each with 10000 runs per value of $\xi\in\{4,5,6,7\}$, we obtain the three results for the optimal route shown in Tab. \ref{tab:results}. 

\begin{table}[h]
    \centering
    \caption{Optimal routes according to each of the 10000-run experiments performed. The possible number of steps  (without counting the starting and finishing points Hobbiton) was $\xi \in\{4,5,6,7\}$, and some of the answers have 5 and some others have 6. For each experiment, all results with higher total priority but total time higher than $t_{MAX}=100$ days were discarded. Shown runtime represents the sum of both the classical and the quantum parts of the process.}
    \begin{tabular}{|c|c|c|c|}
    \hline
    ROUTE & PRIORITY & TRAVEL TIME & RUNTIME \\
    \hline
    Hobbiton &  & & \\
    Valle    &  & & \\
    Isengard   &  & & \\
    Edoras    & 495 & 99 days & 117.57s\\
    Pelargir &  & & \\
    Minas Tirith &  & & \\
    Lórien &  & & \\
    Hobbiton &  & & \\
    \hline
    Hobbiton &  & & \\
    Lórien    &  & & \\
    Pelargir   &  & & \\
    Minas Tirith    & 480 & 96 days & 157.92s\\
    Edoras &  & & \\
    Isengard &  & & \\
    Hobbiton &  & & \\
    \hline
    Hobbiton &  & & \\
    Bree    &  & & \\
    Lórien   &  & & \\
    Minas Tirith    & 460 & 95 days & 131.05s\\
    Pelargir &  & & \\
    Edoras &  & & \\
    Hobbiton &  & & \\
    \hline
    \end{tabular}
    \label{tab:results}
\end{table}

\begin{figure}[h]
  \includegraphics[width=\linewidth]{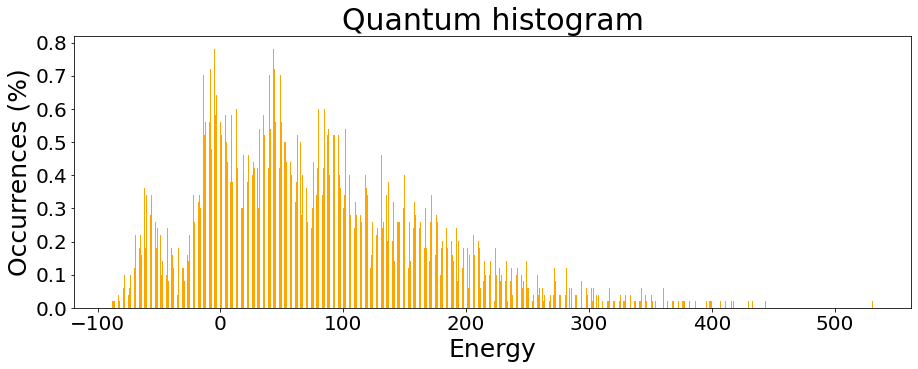}
  \caption{Distribution of the energies of the answers of the quantum resolution of the JSP problem. The first hill, in the negative energies, corresponds to the answers represented in \ref{fig:exhaustive histo}. The rest of the histogram is composed by results with $H^R>0$.}
  \label{fig:quantum histo}
\end{figure}

Only once in these three experiments, $30000$ runs, we obtain one of the $p=495$ optimal solutions. Being $\frac{9!}{(9-6)!}=60480$ different $\xi=6$ routes, with 12 of them having $p=495$, it looks like the quantum solution is similar to a classical random guess. {However, that direct comparison is not entirely symmetrical, since the search space of the quantum device with $N=9$ places and $\xi=6$ steps has $N\xi$ variables, which create a space inhabited by $2^{N\xi}\approx1.8\times10^{16}$ solutions. The distribution of the quantum histogram is shown in Fig. \ref{fig:quantum histo}}.

{For a more solid statistical analysis more data should be gathered, however, only limited access to the device was available at the time of the experiments. Moreover, due to those limitations $10^4$ was the maximum number of runs we could ask for each experiment. If in our results at least one of $30000$ turned out to be an optimal answer, for an increase of the number of runs of only an order of magnitude, up to $10^5$, a sufficient reliability is to be expected.}

\section{\label{sec:conclusions}Conclusions}

The present work manages to solve the proposed Job Selection Problem using the Advantage$\_$system4.1 DWave quantum annealer. Even if it does not achieve any advantage over classical methods, that is due to the size difference in the search space of the two approaches. Improved physical devices which accumulate probability around low-energy answers with greater efficiency would allow the method to be directly implementable.

{Moreover, as shown in Sec. \ref{sec:hamil}, our method only uses $N\xi$ variables ---which also means using $N\xi$  qubits- to solve the problem, as opposed to the $N^2$ variables needed by other methods devised with only the TSP in mind.} This is not only is a notable improvement in resource economy, but a decrease of the answer space size, which we have shown is crucial to take advantage of the probability increase of low-energy states and reliably find the optimum answers.

{Further lines of research coming out of this work include a search for a standarized form of coefficient selection that minimizes the relative energy of the ground state, since is this energy what correlates with the probability of measuring that optimal answer. On the other hand, a standard test of benchmarking the quality of quantum devices could be constructed by evaluating the shape of probability distributions such as the one shown in Figs. \ref{fig:equiprobable} and \ref{fig:quantum histo} for simple, controllable problems similar to the one explained in Sec. \ref{sec:advantage}. Finally, comparison with classical methods more refined than the exhaustive or the equiprobably random used in this paper would also be of interest.}

\medskip

\section*{\label{sec:aknowledgements}Acknowledgments}
The research leading to this paper has received funding from the QUANTEK project (ELKARTEK program from the Basque Government, no. KK-2021/00070).

© 2022 IEEE. Personal use of this material is permitted. Permission from IEEE must be obtained for all other uses, in any current or future media, including reprinting/republishing this material for advertising or promotional purposes, creating new collective works, for resale or redistribution to servers or lists, or reuse of any copyrighted component of this work in other works.

\section*{\label{sec:interests}Competing interests}
The authors declare no competing interests. We acknowledge use of the DWave for this work. The views expressed are those of the authors and do not reflect the official policy or position of DWave or the DWave team.

\bibliographystyle{IEEEtran}
\bibliography{apssamp}

\end{document}